\documentclass[aps,prd,showpacs,preprint,amsfonts]{revtex4}
\usepackage{epsfig}
\usepackage{graphics}
\usepackage{amsmath}
\newcommand{\Eq}[1]{Eq.~(\ref{#1})}
\newcommand{\Eqs}[1]{Eqs.~(\ref{#1})}

\newcommand{\be}{\begin{equation}}
\newcommand{\ee}{\end{equation}}
\newcommand{\bea}{\begin{eqnarray}}
\newcommand{\eea}{\end{eqnarray}}

\newcommand{\pd}[1]{\partial_{#1}}
\newcommand{\pu}[1]{\partial^{#1}}
\newcommand{\EA}[2]{E_{A{#1}}^{#2}}

\newcommand{\Lag}{\mathcal{L}}
\newcommand{\sqrdet}{(-g)^{1/2}}

\def\half{{\scriptstyle{\frac{1}{2}}}}
\def\quarter{{\scriptstyle{\frac{1}{4}}}}
\def\threehalf{{\scriptstyle{\frac{3}{2}}}}

\begin{document}

\author{Jacob D. Bekenstein\footnote{Electronic address: bekenste@phys.huji.ac.il} and Bibhas Ranjan Majhi}\email{bibhas.majhi@mail.huji.ac.il} 
\affiliation{Racah Institute of Physics, Hebrew University of
Jerusalem, Jerusalem 91904, Israel\\}\date{\today}
\title{Is the  principle of least action a must?}
 

\begin{abstract}
The least action principle occupies a central part in contemporary physics.  Yet, as far as classical field theory is concerned, it may not be as essential as generally thought.  We show with three detailed examples of classical interacting field theories that it is possible, in cases of physical interest, to derive the correct field equations for all fields from the action (which we regard as defining the theory), some of its symmetries, and the conservation law of energy-momentum (this last regarded as ultimately coming from experiment).

\end{abstract}
\maketitle

\section{Introduction}
\label{sec:intro}

The principle of least action or Hamilton's principle (henceforth LAP) occupies a central position in contemporary physics.  Beginning with Lagrange and Euler's particle dynamics, continuing through field theory and culminating with string theory, the equations of motion of a \emph{fundamental} theory are generally derived by starting from the theory's action and then invoking the LAP.   Another central pillar of physics is the N\" other theorem whereby a continuous symmetry of a field's action, when combined with the field equations (themselves typically derived from LAP), yields the conservation law (continuity equation) for some continuous field quantity~\cite{Weinberg}.

In simple cases the above logic can be inverted.  A well known example is furnished by the theory of the pure (sourceless) electromagnetic field.  One may infer its action from the requirements of gauge symmetry, time reversal symmetry and Lorentz invariance.   From that action one infers the energy momentum tensor (by the well known recipes for the canonical or symmetric one).  Then by requiring that the tensor's divergence vanish (energy-momentum conservation) one obtains a set of equations which, for generic values of the Faraday tensor, imply that the latter's divergence must vanish~\cite{MTW}.  Thus the Gauss and Ampere laws emerge from energy-momentum conservation, itself a consequence of the symmetry of the action under spacetime translations.  The magnetic Gauss and Faraday laws are automatic consequences of the use of the electromagnetic 4-potential as basic field variable.  Thus, in the mentioned example, the symmetries plus energy-momentum conservation  lead to the field equations without any appeal being made to the LAP.

Other examples are almost as easy to implement when they deal with a single field theory (including a self-interacting one), or with one describing several noninteracting fields.  In the later case separate conservation laws for energy-momentum exist for each field, and the procedure very much follows that for a single field.  However, once different fields are coupled, a single conservation law of each type exists for the whole set of fields, and obviously supplies considerably less information than in the noninteracting case.  Can the field equations for the separate fields still be inferred without appealing to the LAP?

It is germane to mention here the case of dissipative fluid mechanics. Today we tend to think of hydrodynamics as the outcome of integrating out certain degrees of freedom of an underlying fundamental interacting field theory which describes the relevant particles.  It is well known that the energy momentum tensor of the fluid can be written without reference to the underlying fundamental theory (with certain dissipative coefficients appearing as free parameters).  Thereafter the implementation of the laws of particle numbers and energy-momentum conservation will yield the Navier-Stokes equation of motion.  This case suggests it might be possible to recover the various field equations for an interacting field theory from energy-momentum conservation without invoking the LAP.  Is this the case?

This question is interesting inasmuch as it is easily transmuted into a more fundamental one.  From a practical point of view one can hardly analyze a field theory, or predict with it, without knowing its field equations.   But which are the primary entities essential in reaching the field equation? Is it the Lagrangian and LAP, or is it symmetries, the action and conservation laws? 

In this paper we present three nontrivial examples of interacting field systems; for each we recover the correct field equations invoking some of the symmetries and overall energy-momentum conservation, but never recurring to the LAP.  We thus suspect that in large tracts of the field theory zoo, all field equations can be recovered from symmetries and conservation laws when judiciously applied to the action suggested by symmetries, while dispensing with the LAP. Symmetries are, of course, an intuitive primitive concept in physics.  And it is well known how to go about formulating an action for a set of fields with definite content from the requirement that the action incorporate the symmetries.  In addition we can think (and in our approach we \textit{should think}) of conservation laws as experimental facts.  After all the experimental verification of a conservation law is in a real sense a null experiment, and thus very accurate by nature.  

But lest a misunderstanding occur we hasten to add that, as a matter of practice, the traditional way to the field equations via LAP is the short and easy way.  The approach via symmetries and conservation laws described here usually involves somewhat intricate arguments which vary with the nature of the fields, and would not typically be more economical than its traditional counterpart.

Of course all that has been said concerns classical physics.  In quantum field theory the dynamics comes from Feynman's functional (or path) integral.  The exponential in its integrand contains the field's action, so that the starting point of a quantum field theory is similar to that of its classical counterpart.  Since the functional integral cannot usually be calculated exactly, one opts for an approximation scheme such as the popular loop approximation scheme.  In lowest (semiclassical or tree) approximation, this scheme entails evaluating the said exponential at the extremal value of the action with initial and final conditions for the fields being specified.  But this extremal value is, of course, determined by the LAP, and obtains precisely for a field configuration that evolves according to the classical field equations.  Thus LAP has a natural place in the so called semiclassical approximation for the functional integral. However, we are here concerned solely with classical field theory.  This subject can be regarded as autonomous, as witness the large number of texts that develop it independently of quantum considerations. 

The plan of the paper is as follows.  In Sec.~\ref{sec:scalar} we study the most general system of scalar fields in curved spacetime with minimal coupling to the metric.  We show that in $d$-dimensional spacetime the scalar field equations for $d$ or fewer fields scan be derived without help of the LAP.  In Sec.~\ref{sec:electro} we consider a massive charged scalar field interacting with the electromagnetic field.  We demonstrate in flat spacetime that the scalar and electromagnetic field equations can be obtained from symmetry and conservation considerations without use of the LAP.  In Sec.~\ref{sec:T-tau} we take up the theory of Dirac fields interacting with a non-Abelian gauge field, typically with symmetry of $SU(n)$.  We show that the gauge field equation emerges without use of the LAP; likewise, for a representation whose dimension is no higher than half the number of generators minus one [$n^2-2$ for $SU(n)$], the Dirac-like field equation may be obtained from symmetry and conservation considerations alone.

We use natural units in which $\hbar=c=1$.  We employ the metric signature $(-,+,+,+,\dots)$.  Latin indices $a,b,c,d,k$ are spacetime indices, while Greek indices $\mu,\nu,\lambda, \sigma$ denote internal (group) indices.  Repeated indices in one term are summed over.

\section{Mutually interacting nonlinear scalar fields}
\label{sec:scalar}
\subsection{Lagrangian and definitions}
\label{sec:ScalarLagrangian}

In $d$ dimensional curved spacetime let us take the action as a local, very general one, for a collection of interacting  scalar fields $\phi_\mu$ with  $\mu=1,2,\cdots,N\leq d$:
\be
S=\int \Lag(g^{ab},\phi_1,\cdots,\phi_N,\pd{a}\phi_1,\cdots,\pd{a}\phi_N)\,\sqrdet\,d^d x.
\label{scalaraction}
\ee
A trivial example would be one for which $\Lag$ is a linear combination of invariants of the form $\pd{a}\phi_\mu\pu{a}\phi_\mu$ and $\phi_\mu^2$ for several (noninteracting) scalar fields labelled by index $\mu$.  A somewhat more complex example would be one for two massive interacting scalar fields
\be
\Lag=-\half g^{ab}\pd{a}\phi_1\pd{b}\phi_1-\half m_1^2\phi_1^2-\half g^{ab}\pd{a}\phi_2\pd{b}\phi_2-\half m_2^2\phi_2^2+K(\phi_1^2-\phi_2^2)^2\,.
\ee
But, of course, the action (\ref{scalaraction}) includes cases in which the Lagrangian density is not separable into kinetic and potential parts.
 
In terms of the definition of the functional derivative of the action we introduce the notation
\be
E_{\phi}\equiv \frac{1}{\sqrdet} \frac{\delta S}{\delta\phi}= \frac{\partial\Lag}{\partial\phi}-\frac{1}{\sqrdet}\,\pd{a}\left(\frac{\partial\Lag\sqrdet}{\partial(\pd{a}\phi)}\right)\,,
\label{EuLag}
\ee
as well as
\be 
T_{ab}\equiv -\frac{2}{\sqrdet}\frac{\delta S}{\delta g^{ab}}=-\frac{2}{\sqrdet}\frac{\partial \mathcal{L}\sqrdet}{\partial g^{ab}}+\cdots\,.
\label{Tmunu}
\ee
Definition (\ref{Tmunu}) does not necessarily imply that we are using the LAP.  As well known~\cite{LLFields}, it emerges just as a consequence of the diffeomorphism invariance of the non-gravitational part of the action, in which one regards the metric as given, and does not enter into the question of which equations determine it.   We shall employ the definition (\ref{Tmunu}) throughout.
  
\subsection{Consequences of diffeomorphism invariance}
\label{sec:diffeo}
We now consider an infinitesimal increment of the action, $\delta S$, engendered by infinitesimal increments of $\phi_1,\cdots,\phi_N$ in the $d$ dimensional volume $\mathcal{V}$:
\be
\delta S = \int_\mathcal{V} (-\half T_{ab}\,\delta g^{ab}+E_{\phi_1}\delta\phi_1+\cdots+E_{\phi_N}\delta\phi_N)\,\sqrdet d^d x\, +\, \oint_{\partial \mathcal{V}} \left(\frac{\partial\Lag}{\partial(\pd{a}\phi_\mu)}\,\delta\phi_\mu\,\sqrdet\right)\,d\Sigma_a,
\ee
where we sum over $\mu=1,\cdots,N$, and have used Gauss' theorem  to convert a divergence into an integral over the $d-1$ dimensional boundary $\partial\mathcal{V}$ with volume element $d\Sigma_a$.

At this stage we part ways with the usual derivation via the LAP.  We shall only assume that $S$ is invariant under diffeomorphisms $x^a\to x^a+\chi^a(x)$ with the $\chi^a$ being $d$ arbitrary differentiable functions within $\mathcal{V}$, but vanishing on $\partial\mathcal{V}$ and outside $\mathcal{V}$.  Under an infinitesimal  diffeomorphism of this kind, implemented as a Lie drag~\cite{Carroll,Schutz}, $\delta_{\chi}\phi=\chi^a\pd{a}\phi$ as well as $\delta_\chi g^{ab}=-(\chi^{a;b}+\chi^{b;a})$.  Thus the boundary term vanishes and we can write
\be
\delta_\chi S = \int_\mathcal{V} (-T_{ab}{}^{;b}+E_{\phi_1}\pd{a}\phi_1+\cdots+E_{\phi_N}\pd{a}\phi_N)\,\chi^a\,\sqrdet\,d^d x=0\,, 
\ee
where we have exploited the symmetry of $T_{ab}$ and integrated by parts by virtue of the identity
\be
\int_\mathcal{V} T_{ab}\,\chi^{a;b}\sqrdet\,d^d x=-\int_\mathcal{V} T_{ab}{}^{;b}\,\chi^a\sqrdet\,d^d x+\oint_{\partial\mathcal{V}} (T_a{}^b\,\chi^a)\,d\Sigma_b\,,
\ee
and exploited the fact that $\chi^a$ vanishes on $\partial\mathcal{V}$.  Once we take into account the conservation of energy-momentum, $T_{ab}{}^{;b}=0$,  we obtain
\be
\int_\mathcal{V} (E_{\phi_1}\pd{a}\phi_1+\cdots+E_{\phi_N}\pd{a}\phi_N)\,\chi^a\,\sqrdet\,d^d x=0\,. 
\ee
The arbitrariness of the $\chi^a$ allows us to conclude that within $\mathcal{V}$ for every index $a$
\be
 E_{\phi_1}\pd{a}\phi_1+\cdots+E_{\phi_N}\pd{a}\phi_N=0.
 \label{res1}
\ee

For a single scalar field $\phi$ this immediately implies that $E_{\phi}=0$ since in the generic situation $\pd{a}\phi_1\neq 0$ throughout $\mathcal{V}$ apart, perhaps, from special points or surfaces.  By continuity $E_{\phi}=0$ also at these special locations.  Thus we see from \Eq{EuLag} that the field obeys the usual field equation, and this without the LAP having been invoked.  How to extend this argument to many fields? 

\subsection{The argument for several fields}
\label{sec:several}

Let us select $N$ different coordinate components of \Eq{res1} and label them by $a=k_1,k_2,\cdots. k_N$; each component consists of $N$ terms.   The entire set can be written in terms of matrices as
\be
\begin{pmatrix}
\pd{k_1}\phi_1\ \ \pd{k_1}\phi_2\  \cdots\ \pd{k_1}\phi_N\\
\pd{k_2}\phi_1\ \ \pd{k_2}\phi_2\  \cdots\ \pd{k_2}\phi_N\\
\vdots\cr
\pd{k_N}\phi_1\ \ \pd{k_N}\phi_2\  \cdots\ \pd{k_N}\phi_N\\
\end{pmatrix}    \begin{pmatrix}   E_{\phi_1} \\  E_{\phi_2} \\
 \vdots  \\   E_{\phi_N} \\
\end{pmatrix}=0.
\label{matrix}
\ee
In this and the next three paragraph we regard those coordinates $x^a$ that are not included in $\{x^{k_1},x^{k_2},\cdots,x^{k_N}\}$ as fixed; thus we at first consider only a subspace $\mathcal{V}$ of the whole spacetime  spanned by the $\{x^{k_1},x^{k_2},\cdots,x^{k_N}\}$.  This only coincides with the whole spacetime when $N=d$. 

The determinant of the square matrix here is the Jacobian
\be
J=\frac{\partial(\phi_1, \phi_2,\cdots,\phi_N)}{\partial(x^{k_1},x^{k_2},\cdots,x^{k_N})}\,.
\ee
Generically $J$ cannot vanish.  For were it to vanish this would indicate that, viewed as functions of $x^{k_1},x^{k_2},\cdots,x^{k_N}$, the $\phi_1, \phi_2,\cdots,\phi_N$ are functionally related.  This, of course, will not be true in the generic physical field configuration.  It can thus be assumed that $J\neq 0$, except perhaps at isolated point sets.  Then the nonvanishing character of $J$ tells us that the $E_\phi$ column vector must vanish (including at the special points by the argument of continuity).  But the vanishing of this ``vector'' is just the collection of all the Euler-Lagrange equations, \Eqs{EuLag}, for the set of fields $\phi_\mu$.  These field equations are applicable within the said subspace of spacetime in $\mathcal{V}$. 

As mentioned earlier, if $N=d$, $\mathcal{V}$ covers the whole spacetime.  If $N=d-1$ we can set up $d$ distinct equations of the form (\ref{matrix}), one for each coordinate left out the list $\{x^{k_1},x^{k_2},\cdots,x^{k_d}\}$. Carrying out the above procedure using the $d$ distinct Jacobians allows us to extend the previous conclusion to the whole of spacetime when $\mathcal{V}$ itself is allowed to expand without bound.  The above strategy can be suitably generalized for $N\leq d-2$.  We may thus obtain all the Euler-Lagrange field equations all over spacetime, and this without appeal to the LAP. 

Of course, if the number of fields $\phi_\mu$ exceeds $d$, the above analysis, by itself, is insufficient to obtain all field equations.  One would then have to appeal to other symmetries.

\section{Scalar electrodynamics}
\label{sec:electro}

\subsection{Gauge invariant action}
\label{sec:gaugeLag}
 
Here the theory representing a charged (complex) scalar field interacting with a $SU(1)$ gauge field will be considered~\cite{Huang}. As we saw in Sec.~\ref{sec:diffeo} the principal effect of curved space is to make a term containing the energy-momentum tensor appear under an integral when we carry out a diffeomorphism.  To save labour we presume that term has been dropped by invoking energy conservation, as we did earlier.  Thus we may revert to Minkowski spacetime.  We also restrict attention to four spacetime dimensions. The extension to higher dimensional flat spacetime is straightforward.    
  
The action on the four dimensional Minkowski background is given by       
\be
S =\int \mathcal{L}\, d^4x=  \int  \Big[-\frac{1}{4} F^{ab} F_{ab} - D_a\phi D^a{}^* \phi^*  -m^2 \phi^*\phi \Big]   \, d^4x~.
\label{U(1)}
\ee
where the gauge covariant derivative for scalars is defined by $D_a\equiv \partial_a+\imath e A_a$, and the field strength by $F_{ab}=\partial_a A_b -\partial_b A_a$. Here $A_a$ is the gauge vector potential and $e$ is the coupling constant.  This action is gauge invariant under local gauge transformations $A_a\to A_a+e^{-1}\partial_a \Lambda$, $\phi\to \phi \exp(-\imath \Lambda)$  and $\phi^*\to \phi^* \exp(\imath  \Lambda)$, where $\Lambda$, the gauge function, is an arbitrary function of spacetime.  
In analogy with \Eq{EuLag}, denote the functional derivatives of the action with respect to the fields  by the following symbols: 
\be
E_A^a=\frac{\delta S}{\delta A_a}, \qquad E_\phi=\frac{\delta S}{\delta \phi}, \qquad E_{\phi^*}=\frac{\delta S}{\delta \phi^*}\,.
\label{Euler}
\ee

We could have added a potential $V(\phi^*\phi)$ to $\mathcal{L}$; the methods to be described below are easily extended to this case if $V$ is a polynomial in its argument.

\subsection{Scale symmetry and Bianchi identity}
\label{sec:scale}

Now we compute the change engendered in $S$ by any specified infinitesimal increments in $A_a,\phi$ and $\phi^*$.  After integrating by parts and discarding the boundary terms (the last requires that $\delta\phi, \delta\phi^*$ and $\delta A_a$ to vanish on the boundary), as done in Sec.~\ref{sec:scalar},  we get
\be
\delta S = \int ( E_A^a \delta A_a + E_\phi \delta \phi +  E_{\phi ^*}\delta \phi^* )\,d^4x.
\label{increment}
\ee
In case one also increments \textit{parameters} in the action, extra terms will appear here.

It must be noted that apart from gauge symmetry the action \Eq{U(1)} is invariant under a particular kind of scale transformations as well as under diffeomorphisms. This information will now be put to use to find the field equations for the gauge and scalar fields.    

It is easy to check that $S$ is invariant under the scale transformation $x^a\to \varepsilon x^a$, $A_a\to A_a/\varepsilon $, $\phi\to \phi/\varepsilon $, $\phi^*\to  \phi^*/\varepsilon $ and $m\to m/\varepsilon$, with $\varepsilon$ constant throughout spacetime.  The metric is here regarded as scale invariant (the relevant scaling changes being taken up by the coordinates).   We now interpret the collection of small increments mentioned earlier as due to a small scale transformation; accordinglys$\varepsilon=1+\delta\varepsilon$ with $\delta\varepsilon$ infinitesimal. We thus have $\delta_\varepsilon A_a=-\delta\varepsilon A_a, \delta_\varepsilon \phi=-\delta\varepsilon\, \phi$, $\delta_\varepsilon\phi^*=-\delta\varepsilon\, \phi^*$ as well as $\delta_\varepsilon m= -\delta\varepsilon\, m$. Also, the volume element changes as $\delta_\varepsilon d^4x = 4\,\delta\varepsilon\, d^4x$.  In view of all these we obtain to $\mathcal{O}(\varepsilon)$
\be
\delta_\varepsilon S = -\delta\varepsilon \int (  E_A^a  A_a + E_\phi  \phi +  E_{\phi ^*} \phi^* - 2m^2\phi\phi^* -4 \mathcal{L})\, d^4x =0\,.
\label{deps}
\ee
The boundary terms in this particular case are annulled provided that  $A_a,\phi$ and $\phi^*$ themselves vanish on the boundary.  This is certainly implementable if the boundary is taken to be at infinity.

Of course, since $S$ is scale invariant in any gauge, \Eq{deps} should hold in any gauge provided $\varepsilon$ is the same in all gauges.  Now by evaluating the explicit expressions from \Eq{Euler} it may be verified that $E_A^a, E_\phi \phi$ and $E_{\phi^*}\phi^*$, as well as $\phi\phi^*$ are all gauge invariant.  Likewise $\mathcal{L}$ from \Eq{U(1)} is gauge invariant.  But, as we know, the vector potential is not: $A_a\to A_a+e^{-1}\partial_a \Lambda$.  Thus we obtain
\begin{equation}
\delta_\Lambda(\delta_\varepsilon S) = -\delta\varepsilon \int e^{-1}E_A^a\, \partial_a \Lambda\, d^4x = -\delta\varepsilon \int e^{-1}\Big[ \partial_a( E_A^a  \Lambda) - \Lambda\partial_a E^a_A\Big]\, d^4x\,.
\end{equation} 
Now requiring that $\delta_\Lambda(\delta_\varepsilon S) =0$,  discarding the boundary term (which entails requiring $\Lambda=0$ on the boundary) and taking into account the arbitrariness of $\Lambda$ everywhere else, we get
\be
\partial_a E_A^a=0.
\label{EA}
\ee
With the benefit of hindsight this can be recognized as a combination of a trivial identity and the conservation of charge, namely $\pd{a}(\pd{b}F^{ab}-J^a)=0$, where $J^a$ is the $U(1)$ charge current.

\subsection{Gauge symmetry and charge conservation}
\label{sec:gauge-charge}

Let us now go back to \Eq{increment} and interpret the increments mentioned there as due to an infinitesimal gauge transformation: $\delta_\Lambda A_a=e^{-1}\partial_a \Lambda$, $\delta_\Lambda\phi=-\imath\Lambda\phi$  and $\delta_\Lambda\phi^*= \imath \Lambda\phi^*$.   Discarding the boundary terms entails here having both $\Lambda$ and $\pd{a}\Lambda$ vanish on the boundary; we get
\be
\delta_\Lambda S = \int  \Big(e^{-1}\partial_a E^a_A- \imath \phi E_\phi + \imath \phi^* E_{\phi^*} \Big)\,\Lambda\, d^4x=0\,.
\ee
Since $\Lambda$ is arbitrary inside the boundary we find that everywhere in the bulk
\be
\partial_a E_A^a- \imath e \phi E_\phi + \imath e \phi^* E_{\phi^*} =0.
\ee
In view of \Eq{EA}, and the obvious fact that in a generic configuration $\phi$ cannot vanish identically, we see that
\be
\phi E_\phi=\phi^* E_{\phi^*}.
\label{Ephi}
\ee

To elucidate the physical content of this result let us substitute in it the explicit expressions of $E_\phi$ and $E_{\phi^*}$. From \Eq {U(1)} we obtain
\begin{equation}
E_\phi = (D_a D^a\phi)^* - m^2\phi^*; \,\,\,\ E_{\phi^*} = D_a D^a\phi-m^2\phi.
\end{equation}
After substitution in \Eq{Ephi} and some manipulations we obtain 
\be
\partial_aJ^a = 0; \,\,\,\ J^a =\imath e [\phi^*D^a\phi -\phi (D^a\phi)^*]\,,
\label{curre}
\ee
which is recognized as the continuity equation for the conservation of $U(1)$ charge together with the traditional expression for the current.
The factor $\imath e$ in \Eq{curre} has been put in by hand to make the current of the correct dimensions and Hermitian.  It is remarkable that in contrast to the textbook approach, these results  are here obtained without any reference to the explicit equations of motion. The analysis here is ``off-shell''.  

We shall now investigate consequences of the coordinate invariance of the action (\ref{U(1)}).  Actually a full investigation in this direction would have to be performed in curved spacetime since the metric changes in a nontrivial way under diffeomorphisms.  As in Sec.~\ref{sec:diffeo} such change engenders a term proportional to the divergence of the energy momentum tensor.  Since our approach is to impose energy-momentum conservation, this term will drop out.  Hence we have cut corners here, and have done all the work in flat spacetime.  

\subsection{Diffeomorphism symmetry leads to field equations}
\label{sec:diffeo-eq}

We shall now interpret the increments referred to in \Eq{increment} as due to an infinitesimal diffeomorphism, namely $x^a\to x^a + \chi^a$ with $\chi^a$ small.  As in Sec.~\ref{sec:diffeo} we implement this by Lie dragging~\cite{Carroll,Schutz}, to wit  $\delta_{\chi}\phi=\phi,_b \chi^b$; $\delta_\chi\phi^*=\phi^*,_b \chi^b$ and $\delta_\chi A_{a}=\chi^b\partial_b A_a+ A_b\,\partial_a \chi^b$s.
Substituting these together with \Eq{Ephi} into \Eq{increment} gives
\be
\delta_\chi S=\int\Big[E_A^a\Big(\chi^b\,\partial_b A_a+ A_b\,\partial_a \chi^b\Big)+ (E_\phi/\phi^*)\Big(\phi^*\phi,_b + \phi \phi^*,_b\Big)\chi^b \Big] d^4x=0\,.
\ee
By calling on \Eq{EA} we may complete the derivative implied by the first term here to get
\be
\int\Big[ \partial_a\Big(E^a_A\,\chi^bA_b\Big) + \Big\{E_A^a F_{ba} + (E_\phi/\phi^*)(\phi^*\phi,_b + \phi \phi^*,_b)\Big\}\chi^b\Big] d^4x=0\,.
\ee
If we agree that the $\chi^a$ all vanish at the boundary, the perfect divergence here in seen to integrate to a vanishing boundary term.  In view of the arbitrariness of $\chi^b$ within $\mathcal{V}$ we get
\be
E_A^a F_{ba} + (E_\phi/\phi^*)\, \pd{b}|\phi|^2=0\,.
\label{EAphi}
\ee
In view of the antisymmetry of $F_{ba}$ contraction of this equation with  $E_A^b$ yields
\be
E_\phi\, E_A^b\, \pd{b}|\phi|^2 =0\,.
\ee

If we ignore the non-generic configuration with $|\phi|=\textrm{const.}$, the above equation has two solutions: either (I) $E_\phi=0$ or (II) $E_A^b\, \pd{b}|\phi|^2=0$.  We consider each case separately. 
\vskip 5mm

\noindent
{\underline{Case I}:  $E_\phi=0$, at least over a finite spacetime region.\\
Of course, since $\phi$ (and consequently $\phi^*$)  cannot vanish identically,  \Eq{Ephi} immediately implies that $E_{\phi^*}=0$ throughout the same region.    It now follows from \Eq{EAphi} that everywhere in that region 
\be
E_A^a F_{ba}=0.
\ee
Now if we write out the determinant of the matrix made up of components $F_{ba}$ we see that it cannot vanish identically (its vanishing would imply a very non-generic electromagnetic configuration).  The only solution of the linear system of $d$ equations is thus $E_A^a=0$.   Thus, without invoking the LAP, we have found all the equations of motion in the said spacetime region: $E_A^a=E_\phi=E_{\phi^*}=0$.

\vskip 3mm
\noindent
\underline{Case II}: $E_A^b\, \pd{b}|\phi|^2 =0$.\\
For the generic configuration we cannot have $|\phi|^2$ a spacetime constant.  Hence $E_A^a$ is orthogonal to  $\pd{b}|\phi|^2$, except perhaps on special surfaces or points.  In \textit{curvilinear}  coordinates \Eq{EA} can be written more explicitly as
\be
\pd{a}[\sqrdet E_A^a]=0.
\label{div0}
\ee
The orthogonality of $E_A^a$ and $\pd{b}|\phi|^2$ means that $E_A^a$ lies on constant $|\phi|^2$ hypersurfaces. Thus by choosing one of the coordinates (not necessarily a timelike one), say $x^0$, to coincide with $|\phi|^2$ itself, we get that $E_A^0=0$ throughout.  While this arrangement may not be valid globally, it can certainly be employed separately in suitable patches that cover all spacetime.  

In this essentially 3-D situation we may infer from \Eq{div0} that (a divergenceless vector is necessarily a curl)
\be
\sqrdet E^k=\varepsilon^{klm} \pd{m}W_l\,,
\label{curl}
\ee
where $k,l,m$ label the coordinates other than $x^0$, $W_l$ is some covariant  3-vector, and $\varepsilon^{klm}$ is the 3-D Levi-Civita alternating symbol.  Because $\pd{b}|\phi|^2=\delta_b^0$ we may now rewrite \Eq{curl} in 4-D language as
\be
\sqrdet E_A^a=\pm\varepsilon^{bacd} \pd{d}W_c\,\pd{b}|\phi|^2\,,
\label{curl2}
\ee
where $\varepsilon^{bacd}$ is now the 4-D Levi-Civita alternating symbol, and the sign depends on whether $x^0$ is timeline or not.
This equation is covariant, so we can easily return to the Minkowski version
\be
E_A^a=\pm\varepsilon^{bacd} \pd{d}W_c\,\pd{b}|\phi|^2\,.
\label{curl3}
\ee

We notice that one can add to $W_c$ an arbitrary gradient without thereby affecting $E_A^a$ or the rest of the equations.  This reflects a $U(1)$ symmetry.  Now were the $W_c$  distinct from the electromagnetic potential $A_c$, the theory in question would have a $U(1)\times U(1)$ symmetry.  But such extended symmetry can be associated with electromagnetism only under special circumstances~\cite{BekBet}.  Excluding it here we conclude that necessarily $W_c=K A_c$ with $K$ a constant.

Let us now calculate the left hand side of \Eq{curl3} directly from \Eq{U(1)} using prescription (\ref{Euler}).  Further, by using the antisymmetry of the Levi Civita symbol, we  can the put \Eq{curl3} in the final form
\be
\pd{b}(F^{ab}\pm K\varepsilon^{bacd} \pd{d}A_c|\phi|^2)=J^a\,,
\label{mixed}
\ee
where $J^a$ is defined in \Eq{curre}.  We notice now that the argument of the divergence is a linear combination of a true antisymmetric tensor, $F^{ab}$, and a pseudo-antisymmetric tensor, 
$\varepsilon^{bacd} \pd{d}A_c$.  These two entities transform oppositely under spatial inversion, which means that if $K\neq 0$, the electromagnetic field equation that emerges breaks the spatial inversion symmetry of the Lagrangian in \Eq{U(1)}.  To avoid such unnatural behavior we set $K=0$; it follows that $E_A^a=0$.   Note that the phenomenon encapsulated in \Eq{mixed} is distinct from spontaneous symmetry breaking whereby specific solutions flout a symmetry of the field equations.   It is also different from the appearance of duality symmetry, e.g. in electromagnetism,  because this last one occurs only in the absence of sources, whereas \Eq{mixed} has a current source.
  
Now because in a generic configuration $|\phi|^2$ cannot be a spacetime constant, we can only satisfy \Eq{EAphi} by having $E_\phi=0$.  Of course by \Eq{Ephi} it also follows that $E_{\phi^*}=0$.  Thus we have again found $E_A^a=E_\phi=E_{\phi^*}=0$, i.e. all the field equations for the problem have been established without invoking the LAP.

\section{Fermion interacting with nonabelian gauge field}
\label{sec:T-tau}

\subsection{The gauge theory}
\label{sec:gauge}

The standard model of particle physics abounds with examples of spin-$\half$ fermions interacting via gauge fields.  We thus turn attention to such a case, a spin-$\half$ fermion in interaction with a non-Abelian gauge field.  The action, again formulated in flat 4-D Minkowski spacetime,  is given by~\cite{Weinberg2}
\be
S =\int \mathcal{L} d^4x=  \int  \Big[-\quarter F_\mu^{~ab} F_{\mu ab} + \bar{\psi}(\imath\gamma^a \mathbf{D}_a - m)\psi\Big]   \, d^4x\,.
\label{SU(n)}
\ee
where $\mathbf{D}_a\equiv \partial_a+\imath g A_{a\mu}\mathbf{L}_\mu$ is the gauge covariant derivative for the theory, and $F_{\mu ab}=\partial_a A_{\mu b} -\partial_b A_{\mu a} -g C_{\mu\nu\lambda}A_{\nu b} A_{\lambda a}$. The $C_{\mu\nu\lambda}=-C_{\mu\lambda\nu}$ are the structure constants of whatever group describes the gauge symmetry,  while the $\mathbf{L}_\mu$ are hermitian matrices corresponding to the abstract group generators (a representation of the group). Here $a,b$, etc are spacetime indices while $\mu,\nu$, etc. are group (or color) indices.  That is there are several gauge 4-potentials $A_{\mu a}$ and correspondingly several gauge field tensors $F_{\mu a b}$.

   For the usual theories the $\mathbf{L}_\mu$ satisfy some $SU(n)$ Lie algebra: $[\mathbf{L}_\mu,\mathbf{L}_\nu] = \imath C_{\lambda\mu\nu}\mathbf{L}_{\lambda}$, with the $\mathbf{L}_\mu$, $n^2-1$ in number,  normalized according to  $\mathrm{Tr}(\mathbf{L}_\nu \mathbf{L}_\mu+\mathbf{L}_\mu \mathbf{L}_\nu)=2\delta_{\nu\mu}$; they are of size $N\times N$ when the theory uses a representation of the group of order $N$. Finally the $\psi$ is a multiplet, a column of $N$ Dirac spinors; $\bar\psi$, the adjoint multiplet spinor, is a row containing $N$ adjoint 4-spinors of the form familiar from the theory of the Dirac equation.  The $\mathbf{L}_\mu$ act on the multiplet (or color) space while the usual Dirac $\gamma^a$ matrices act equally on each of the Dirac spinors composing each multiplet.
 
The action (\ref{SU(n)}) is known to be invariant under non-Abelian gauge transformations.  Since we are interested below in how the various field quantities behave under such transformations, we shall go into some detail into how this invariance comes about.   First one constructs the spacetime dependent unitary matrix $\mathbf{U}=\exp[-\imath \mathbf{L}_\mu \Lambda_\mu(x)]$, where the $\Lambda_\mu$ are arbitrary real space-time functions.  Likewise one forms matrix versions of the vector potentials and fields: $\mathbf{A}_a\equiv \mathbf{L}_\mu A_{\mu a}$ and $\mathbf{F}_{ab}\equiv \mathbf{L}_\mu F_{\mu a b}$.  By analogy $\mathbf{D}_a\psi= (\pd{a}+\imath g \mathbf{A}_a)\psi$.   Since the $\mathbf{L}_\mu$ are independent it is easy to recover the individual $A_{\mu a}$ from the $\mathbf{A}_a$, etc.

 The fermion fields are taken to transform as $\psi\to \mathbf{U}\psi$ and $ \bar\psi\to \bar\psi\, \mathbf{U}^\dagger$.
In addition, one prescribes the transformation law 
\be
\mathbf{A}_a \to \mathbf{U} \mathbf{A}_a \mathbf{U}^\dagger + \imath g^{-1} (\pd{a} \mathbf{U})\, \mathbf{U}^\dagger\,,
\label{TransfA}
\ee
as a result of which the field matrix and the covariant derivative transform covariantly, namely
$\mathbf{F}_{ab} \to \mathbf{U} \mathbf{F}_{a b} \mathbf{U}^\dagger$ and   $\mathbf{D}_a\psi\to \mathbf{U} \mathbf{D}_a\psi$.  The action (\ref{SU(n)}) can now be rewritten  as
\be
S=\int \big[-\quarter \mathrm{Tr}(\mathbf{F}_{ab}\mathbf{F}^{ab}) + \bar{\psi}(\imath\gamma^a \mathbf{D}_a - m)\psi\big]   \, d^4x\,. 
\ee
By making the above-mentioned transformations of $\mathbf{F}_{ab}, \mathbf{D}_a, \psi$ and $\bar\psi$ and invoking the cyclic invariance of the trace of a product of matrices it is immediate to see that $S$ and its Lagrangian density are both unchanged, i.e. gauge invariant.

We may denote the variational derivatives with respect to the system's field variables by
\be
E_{A\mu}^a=\frac{\delta S}{\delta A_{\mu a}}, \qquad E_\psi=\frac{\delta S}{\delta \psi}, \qquad E_{\bar{\psi}}=\frac{\delta S}{\delta {\bar{\psi}}}~.
\label{nonvariation}
\ee
Thus to any increment of the fields corresponds the increment of the action 
\be
\delta S = \int ( E_{A\mu}^a \delta A_{\mu a} + E_\psi \delta \psi + \delta\bar{\psi} E_{\bar{\psi}})\, d^4x\,,
\label{nonvaction}
\ee
with
\bea
E_{\bar\psi}&=&(\imath\gamma^a \mathbf{D}_a-m)\psi\,,
\label{E1}
\\
E_\psi&=&- \imath\partial_a\bar\psi \gamma^a -\bar\psi(g\gamma^a A_{\mu a} \mathbf{L}_\mu + m)\,,
\label{E2}
\\
E^a_{A\mu}&=& -\nabla_b F^{ab}_\mu -g\bar{\psi}\gamma^a\mathbf{L}_\mu\psi
\label{E3}
\eea
where $\nabla_b F^{ab}_\mu \equiv \pd{b} F^{ab}_\mu +g\, C_{\sigma\nu\mu}\, A_{\nu b}\, F_\sigma^{ab}$ defines the \textit{gauge covariant divergence} of a tensor.  
As before, we have dropped the boundary terms; this is natural when the boundary lies at infinity and $A_{\nu b}$, $\psi$, $\bar\psi$ and their derivatives vanish asymptotically.
\subsection{Scale and gauge symmetry lead to the gauge field equations}
\label{sec:scale-gauge}

We now observe that the action (\ref{SU(n)}) is also invariant under the scale transformation $x^a\to \varepsilon x^a$, $A_{\mu a}\to A_{\mu a}/\varepsilon $, $\psi\to \varepsilon^{-3/2}\psi $,  $\bar{\psi} \to  \varepsilon^{-3/2}\bar{\psi} $, $m\to m/\varepsilon$ with $\varepsilon$ a positive constant.  The metric is again regarded as unchanged. Under an infinitesimal scale transformation with $\varepsilon=1+\delta\varepsilon$ we have that $\delta_\varepsilon A_{\mu a}=-\delta\varepsilon\,A_{\mu a},\, \delta_{\varepsilon}\, \psi=-\threehalf\delta\varepsilon\, \psi, \,\delta_{\varepsilon}\bar{\psi}=-\threehalf\delta\varepsilon\, \bar{\psi}, \delta_{\varepsilon} m=-\delta\varepsilon\, m$ and   $\delta_\varepsilon d^4x \to 4\delta\varepsilon\, d^4x$.  Hence
\be
\delta_\varepsilon S = -\delta\varepsilon \int \big(  E_{A\mu}^a  A_{\mu a} + \threehalf E_\psi  \psi + \threehalf \bar{\psi} E_{\bar{\psi}}-m\bar\psi\psi - 4 \mathcal{L}\big)\, d^4x=0\,.
\ee
Here we drop boundary terms under the conditions already mentioned.   The above result can be be written as ($\mathbf{E}^a_A=\mathbf{L}_\mu E_{A\mu}^a $)
\be
\int \mathrm{Tr}\, (\mathbf{E}^a_A \mathbf{A}_a)\, d^4x=\int \big(4\mathcal{L}+m\bar\psi\psi-\threehalf
E_\psi \psi-\threehalf \bar\psi E_{\bar\psi}\big) d^4x.
\label{dS}
\ee

Result (\ref{dS}) must hold in any gauge because  $S$ is scale invariant in any gauge if $\varepsilon$ is the same in all gauges.   Now $\bar\psi\psi$ is evidently gauge invariant.  By combining \Eq{E1} with \Eq{TransfA} it is readily shown that $\bar\psi E_{\bar\psi}$ is gauge invariant.  Similarly $E_\psi\psi$ is gauge invariant.  We already know that $\mathcal{L}$ is gauge invariant.   Therefore,  $\int \mathrm{Tr}\, (\mathbf{E}^a_A \mathbf{A}_a)\,d^4x$ must itself be gauge invariant.

Now according to \Eq{E3} $E_{A\mu}^a$ is the sum of the gauge covariant derivative of $F_\mu^{ab}$ and the quantity $-g\bar{\psi}\gamma^a\mathbf{L}_\mu\psi$ which is itself known to be gauge covariant~\cite{Weinberg2}.  This means means that under a gauge transformation $\mathbf{E}^a_A\to \mathbf{U} \mathbf{E}^a_A \mathbf{U}^\dagger$.  Employing \Eq{TransfA} 
we see that
\bea
\mathrm{Tr}\, (\mathbf{E}^a_A \mathbf{A}_a)&\to& \mathrm{Tr}\big[(\mathbf{U}\mathbf{E}^a_A \mathbf{U}^\dagger \mathbf{U} \mathbf{A}_a \mathbf{U}^\dagger)\big]+ \imath g^{-1} \mathrm{Tr}\big[\mathbf{U}\mathbf{E}^a_A \mathbf{U}^\dagger (\pd{a} \mathbf{U}) \mathbf{U}^\dagger\big]
\\
&=& \mathrm{Tr}\, (\mathbf{E}^a_A \mathbf{A}_a)+\imath g^{-1}  \mathrm{Tr}\big[\mathbf{E}^a_A \mathbf{U}^\dagger (\pd{a} \mathbf{U}) \big]
\eea
This tells us immediately that $\int\mathrm{Tr}\big[\mathbf{E}^a_A \mathbf{U}^\dagger (\pd{a} \mathbf{U})\big]\,d^4x=0$.  Now to first order in the $\Lambda_\mu$s, $\mathbf{\mathbf{U}}= 1-\imath \mathbf{L}_\mu \Lambda_\mu+\mathcal{O}(\Lambda^2)$.  Therefore, because $\mathrm{Tr}(\mathbf{L}_\nu \mathbf{L}_\mu+\mathbf{L}_\mu \mathbf{L}_\nu)=2\delta_{\nu\mu}$ we get, correct  to $\mathcal{O}(\Lambda_\mu)$, that
\be
\mathrm{Tr}\big[\mathbf{E}^a_A \mathbf{U}^\dagger (\pd{a} \mathbf{U})\big]=-\imath E^a_{A\mu}\pd{a}\Lambda_\mu=-\imath \pd{a}(E^a_{A\mu}\Lambda_\mu)+\imath \Lambda_\mu \pd{a}E^a_{A\mu}\,.
\ee
Assuming that the $\Lambda_\mu$ vanish asymptotically we get by Gauss theorem that $\int \Lambda_\mu \pd{a}E^a_{A\mu}\,d^4x$ = 0.   But since the $\Lambda_\mu$ are all arbitrary functions, this tells us that for every $\mu$, $\pd{a}E^a_{A\mu}=0$, or equivalently, $\pd{a}\mathbf{E}_A^a=0$ all over spacetime.

Of course this last result must be gauge invariant (we did not choose a specific gauge in deriving it).  Thus for arbitrary $\mathbf{U}$ we should also have  $\pd{a}(\mathbf{U}\mathbf{E}_A^a \mathbf{U}^\dagger)=0$.  Substituting  $\mathbf{U}$ here and retaining terms only up to $\mathcal{O}(\Lambda_\mu)$ we obtain
\be
\pd{a}(\mathbf{U}\mathbf{E}_A^a \mathbf{U}^\dagger)=\pd{a}\mathbf{E}_A^a-\imath \pd{a}\big(E_{A\nu}^a[\mathbf{L}_\mu,\mathbf{L}_\nu]\Lambda_\mu\big)+\mathcal{O}(\Lambda^2),
\ee
so that the term of $\mathcal{O}(\Lambda_\mu)$ must vanish.   
Substituting for the commutator of the $\mathbf{L}_\mu$ and taking into account that the $\mathbf{L}_\mu$ are independent, we infer that for any $\tau$
\be
C_{\tau\mu\nu} \EA{\nu}{a}\pd{a}\Lambda_\mu =0.
\ee

Now, for $SU(n)$ symmetry, the $\Lambda_\mu$ are $n^2-1$ arbitrary functions. At any chosen event the $4(n^2-1)$ quantities $\pd{a}\Lambda_\mu$ can be taken to be arbitrary, and so independent of one another. But then the above mentioned  linear combination of the $4(n^2-1)$ quantities $\EA{\nu}{a}$, whose coefficients involve $4(n^2-1)$ arbitrary degrees of freedom as just mentioned, cannot vanish identically as required unless all the $\EA{\nu}{a}$ themselves do so. This conclusion can be drawn separately for each $\tau$.  According to \Eqs{EuLag} and (\ref{nonvariation}) the conditions $\EA{\nu}{a}=0$ we have just inferred without appealing to the LAP are precisely the gauge field equations for the system, to wit
\be
\nabla_b F^{ab}_\mu =\pd{b} F^{ab}_\mu +g\, C_{\sigma\nu\mu}\, A_{\nu b}\, F_\sigma^{ab}=-g\bar{\psi}\gamma^a \mathbf{L}_\mu\psi\,.
\label{Efieldeq}
\ee

\subsection{Spinor field equation: Abelian case}
\label{sec:U(1)}

We proceed to infer the field equations for the spinor fields. This turns out to be no easy task, so in this subsection we limit discussion to that of a single \textit{Abelian} gauge field interacting with a singlet spin-$\half$ field.  In this case the role of $\mathbf{L}_\nu$ is played by unity, and the multiplet degenerates to a single 4-spinor ($N=1$).

Consider then the (vanishing) increment (\ref{nonvaction}) in the action (\ref{SU(n)}) due to the infinitesimal local gauge transformation $\mathbf{U}= 1-\imath \Lambda+\mathcal{O}(\Lambda^2)$. We substitute  $E_{A}^a=0$ from \Eq{Efieldeq} to get
\be
\delta_\Lambda S =\imath \int(-E_{\psi} \psi+\bar{\psi}\,E_{\bar{\psi}})\Lambda\,d^4x =0\,.
\label{nongauge}
\ee 
But $\Lambda$ is arbitrary, so we get at every spacetime point
\be
E_\psi \psi = \bar\psi E_{\bar\psi}\,.
\label{E4}
\ee
We can substitute in this \Eqs{E1} and (\ref{E2}) to obtain the local conservation law of $U(1)$ charge, e.g. electric charge:
\be
\partial_a (\bar\psi \gamma^a \psi)=0\,.
\label{currcons0}
\ee
This result \textit{appears} to be off-shell since we have not yet identified the spinor field equations.  We now proceed to that task, in which critical use will be made of \Eq{E4}.

We return to the (vanishing) increment, \Eq{nonvaction}, of the action (\ref{SU(n)}), but this time induced by the diffeomorphism  $x^a \rightarrow x^a +\chi^a$, where $\chi^a$ is an infinitesimal but otherwise arbitrary spacetime dependent vector field. First the contribution from the increment in metric vanishes upon enforcement of the local conservation of energy-momentum.  Next we take into account that $E_{A\mu}^a=0$.  Thus we get 
\be
\delta_{\chi}S = \int (E_{\psi}\,\delta_{\chi}\psi + \delta_{\chi}\bar{\psi}\,E_{\bar{\psi}})\,d^4x =0,
\label{dSdiffeo}
\ee 
where the increments in the spinors are produced by Lie dragging them~\cite{Brill&Wheeler}, namely,
\begin{eqnarray}
&&\delta_{\chi}\psi = \chi^a\partial_a\psi + \quarter\imath\,(\partial_a\chi_b)\sigma^{ab}\psi,
\nonumber
\\
&& \delta_{\chi}\bar{\psi} = \chi^a\partial_a\bar{\psi} - \quarter\imath\,\,(\partial_a\chi_b)\bar{\psi}\sigma^{ab}
\end{eqnarray}
with $\sigma^{ab} = \imath/2[\gamma^a,\gamma^b]$.  

Integrating by parts in \Eq{dSdiffeo} and dropping boundary terms as done earlier we obtain
\be
\delta_\chi S = \int d^4x\, \chi^b \big[ E_{\psi}(\partial_b\psi) + (\partial_b\bar\psi)E_{\bar{\psi}} - \quarter\imath\,\partial_a(E_\psi \sigma^{ab}\psi - \bar\psi\sigma^{ab}E_{\bar\psi})\big]=0.
\ee
Thus on account of the arbitrariness of $\chi^b$ we infer that
\be
 E_{\psi}(\partial_b\psi) + (\partial_b\bar{\psi})E_{\bar{\psi}} - \quarter\imath\,\partial_a(E_\psi \sigma^{ab}\psi - \bar{\psi}\sigma^{ab}E_{\bar{\psi}})=0.
 \label{represent}
\ee
Taking the partial derivative $\partial^a$ of both sides and exploiting the antisymmetry of $\sigma^{ab}$ we obtain the simplified relation 
\be
\partial^b[E_\psi\partial_b\psi+(\partial_b\bar\psi) E_{\bar\psi}]=0.
\label{div}
\ee
This looks like a continuity equation; it is gauge covariant as can be verified by replacing $\partial_b\psi\to \partial_b\psi+\imath g A_b\,\psi$ together with its conjugate version, and carrying out a local Abelian gauge transformation. 

However, there is no room for a new conservation law such as \Eq{div}.  It does not correspond to charge conservation which would require a vector current formed without any derivatives of $\psi$. Neither does it correspond to energy-momentum conservation since the quantity whose divergence is zero here is a vector, not a symmetric tensor. The best guess is that the above result, far from being some serendipitous conservation law, is vacuous, i.e., 
\be
V_b\equiv E_\psi\partial_b\psi+(\partial_b\bar\psi) E_{\bar\psi}=0.
\label{Eb}
\ee

Can we  escape this conclusion?  Several possibilities offer themselves.  For example, $V_b$ could be proportional to a Killing vector, which would automatically nullify its divergence as required.  Of course such situation would obtain only in the presence of a spacetime symmetry; for a generic field configuration there would be no Killing vector.  Another possibility would be if all the components of $V^b$ were proportional to $(-g)^{-1/2}$.  This would also nullify the divergence.  However, in such case $V^b(-g)^{1/2}$ would be a 4-vector with constant contravariant components; such a ``constant'' vector---signifying a special spacetime direction--- simply has no place in a generic solution.  We must thus conclude that it is very hard to avoid \Eq{Eb}.

These set of equations is somewhat reminiscent of \Eq{res1} which we used to obtain the field equations for the scalar fields in Sec.~\ref{sec:scalar}.  The difference here is that we have only four equations (one for each of the four coordinates) constraining, so it seems, eight quantities in all, the components of $E_\psi$ and of $E_{\bar\psi}$.  However, we receive assistance from \Eq{E4}.  Suppose we multiply this last by an \textit{arbitrary complex} vector $t_b$ (with no spinor aspects whatsoever), and add the result to \Eq{Eb} to get
\be
E_\psi(\partial_b+t_b)\psi+[(\partial_b-t_b)\bar\psi] E_{\bar\psi}=0.
\label{E5}
\ee
Now we take the Hermitian conjugate of this.  In reworking the result we take into account that Hermitian conjugation inverts the order of all matrices and spinors besides conjugating each.  We further take into account that $\bar\psi\equiv \psi^\dagger \gamma^0$ and that $\gamma^0 (\gamma^a)^\dagger \gamma^0=\gamma^a$ for all $a$ (spatial $\gamma$s are antihermitian).  We thus get
\be
E_\psi(\partial_b-t_b^*)\psi+[(\partial_b+t_b^*)\bar\psi] E_{\bar\psi}=0.
\label{E6}
\ee

In \Eqs{E5} and (\ref{E6}) we have, counting coordinate components, a total of eight linear equations in the eight components of the spinors $E_\psi$ and $E_{\bar\psi}$.  In fact we can write the system as
\be
\left(\begin{array}{cc}(\partial_b-t_b)\bar\psi & (\partial_b+t_b)\psi^T \\(\partial_b+t_b^*)\bar\psi & (\partial_b-t_b^*)\psi^T\end{array}\right)\left(\begin{array}{c} E_{\bar\psi}\\E_\psi{}^T\end{array}\right)=0,
\ee
where ``$T$'' signifies transpose of the relevant spinor.  It is clear that the $8\times 8$ matrix here does not have trivially identical rows or columns for a generic spinor field configuration. For one thing the $t_b$ is an arbitrary set of fields.  And in generic spinor field configuration the relation between $\bar\psi$ and $\psi^T$ will not be a simple one.  Thus the determinant of the matrix can vanish only at isolated points or surfacess.  By continuity this allows us to conclude that in a generic field configuration $E_\psi=E_{\bar\psi}=0$ everywhere in spacetime, i.e. that the spinor field equations are satisfied.  Again, it is obvious that this conclusion did not entail use of the LAP.

\subsection{Gauge symmetry and nonabelian charge conservation}
\label{sec:nonabelian}

Returning now to the nonabelian theory we reconsider the increment (\ref{nonvaction}) in the action (\ref{SU(n)}) due to the infinitesimal local gauge transformation $\mathbf{U}= 1-\imath \mathbf{L}_\mu \Lambda_\mu+\mathcal{O}(\Lambda^2)$. Substituting $E_{A\mu}^a=0$ we get
\be
\delta_\Lambda S =\imath \int(-E_{\psi}\mathbf{L}_\mu \psi+\bar{\psi}\,\mathbf{L}_\mu E_{\bar{\psi}})\Lambda_{\mu}\,d^4x =0\,,
\label{nongauge2}
\ee 
which by virtue of the arbitrariness of the $\Lambda_\mu$ says that
\be
E_\psi \mathbf{L}_\nu \psi = \bar\psi \mathbf{L}_\nu E_{\bar\psi}\,.
\label{E1E2}
\ee
We now substitute \Eqs{E1} and (\ref{E2}) into \Eq{E1E2}.  The mass terms cancel leaving us with
\be
\imath \partial_a (\bar\psi \gamma^a \mathbf{L}_\nu \psi)-gA_{\mu a} \bar\psi \gamma^a (\mathbf{L}_\nu \mathbf{L}_\mu - \mathbf{L}_\mu \mathbf{L}_\nu)\psi=0.
\label{next}
\ee
Employing the group commutation law $\mathbf{L}_\nu \mathbf{L}_\mu - \mathbf{L}_\mu \mathbf{L}_\nu= \imath C_{\lambda\nu\mu}\mathbf{L}_\lambda$ we have
\be
\partial_a J^a_\nu+g\, C_{\lambda\mu\nu}A_{\mu a}J^a_\lambda=0;\quad J_\nu^a\equiv \bar\psi \gamma^a \mathbf{L}_\nu \psi\,.
\label{currcons}
\ee
This is the well known local conservation law for the $SU(n)$ charges (one for each generator $\mathbf{L}_\mu$)~\cite{Huang,Weinberg2}.   The form of \Eq{currcons} parallels that of \Eq{Efieldeq}, which informs us that the former is also gauge covariant. 

As in the abelian case, the above derivation of the charge conservation laws \textit{appears} to off-shell, i.e. we have yet to find the field equations of motion for $\psi$ and $\bar\psi$.

\subsection{A novel partial symmetry}
\label{sec:novel}

While not immediately related to our main line of reasoning, the following discussion turns up a novel symmetry in the system (\ref{SU(n)}).  We observe from \Eq{E1} that the Dirac part of the action (including the gauge interaction term) from \Eq{SU(n)} can be written
\be
S_D=\int \bar\psi E_{\bar\psi} d^4x.
\ee
Consider now the unitary transformation $\psi\to \mathbf{W}\psi$ and $\bar\psi\to \bar\psi\, \mathbf{W}^\dagger$ with
\be
\mathbf{W}=\exp(-\imath\epsilon_\nu \mathbf{L}_\nu)
\ee
where the $\epsilon_\nu$ are real \textit{constants} and we have added to the generators $\mathbf{L}_0=\mathbf{I}$, the unit matrix of the relevant rank. In the $SU(n)$ case there will be $n^2$ different $\epsilon_\nu$s.  The gauge potentials $A_{\nu a}$ are to be regarded as entirely unchanged under the said transformation, which thus differs from a \textit{global}  nonabelian gauge transformation, c.f.~\Eq{TransfA}.  Now because $\mathbf{W}$ commutes with the $\gamma^a$, it is immediate to see that $\bar\psi E_{\bar\psi}\to \bar\psi E_{\bar\psi}+g A_{\mu a} \bar\psi\,\gamma^a (\mathbf{L}_\mu-\mathbf{W}^\dagger \mathbf{L}_\mu \mathbf{W})\psi$.  Expanding $\mathbf{W}$ in a series in the $\epsilon_\mu$ and keeping up to first order terms, we get $\mathbf{L}_\mu-\mathbf{W}^\dagger \mathbf{L}_\mu \mathbf{W}=-\imath\epsilon_\nu [\mathbf{L}_\nu , \mathbf{L}_\mu]= \epsilon_\nu C_{\lambda\nu\mu} \mathbf{L}_\lambda$.  Accordingly under the said transformations
\be
\bar\psi E_{\bar\psi}\to \bar\psi E_{\bar\psi}- g\epsilon_\nu A_{\mu a}  C_{\lambda\mu\nu} \bar\psi\, \gamma^a \mathbf{L}_\lambda\psi=\bar\psi E_{\bar\psi}+\epsilon_\nu \partial_a J^a_\nu,
\ee
where use has been made of \Eq{currcons}.

By Gauss' theorem the spacetime integral of $\partial_a J^a_\mu$ reduces to a boundary term.   Consequently, in essence the total action (\ref{SU(n)})  is invariant under the unitary transformation $\psi\to \mathbf{W}\psi,\,\bar\psi\to \bar\psi\, \mathbf{W}^\dagger,\, A_{\mu a}\to A_{\mu a}$.  By this we mean that the only change is addition of a ``surface term'' which leads, so we know from the standard approach, to no changes in the field equations.  Unlike the usual symmetries, this one requires that we already impose the gauge field equations of motions, \Eq{Efieldeq}, which are the predecessor of the charge conservation laws, \Eq{currcons}.   It must be stressed, however, that no use has been made of the spinor equations of motions which have yet to be formally identified. 

\subsection{Spinor field equations: nonabelian case}
\label{sec:spinor-eq}

Let us now combine all laws of the form (\ref{E1E2}) into the equation (the $\epsilon_\nu$ are, again, real constants)
\be
E_\psi (\epsilon_\nu \mathbf{L}_\nu \psi) - (\bar\psi\epsilon_\nu \mathbf{L}_\nu)\, E_{\bar\psi}=0;\quad 1\leq \nu\leq N.
\label{double}
\ee
We recall that $E_{\psi}$ stands for a row comprising $N$ adjoint 4-spinors $E_{\psi}{}^{(j)}$.  Thus the first term in the equation above is the scalar product of this row with a column with $N$ entries, each of which is a linear superposition of the Dirac 4-spinors $\psi^{(j)}$ that make up $\psi$ (the $\mathbf{L}_\nu$ responsible for the superposition do not mix the individual components of each 4-spinor $\psi^{(j)}$).   The second term is identical to $E_{\bar\psi}{}^T (\epsilon_\nu \mathbf{L}_\nu{}^T \bar\psi^T) $ (again, the ``$T$" here stands for transpose between rows and columns) whose structure is the same as that of the first term, except that the Dirac 4-spinors being superposed are now the $\bar\psi^{(j)}{}^T$.  

Now focus on a specific spinor configuration.  We can think of the $\psi^{(j)}$ and $\bar\psi^{(j)}{}^T$ together as constituting a column of $2N$ Dirac spinors.   Multiplication of it by a $2N\times 2N$ matrix having $\mathbf{L}_\nu$ and $\mathbf{L}_\nu{}^T$ along the diagonal engenders  a ``rotation'' of this column in a space of columns of dimension $2N$.  It is intuitively clear that  multiplying with different $\mathbf{L}_\nu,\ \mathbf{L}_\nu{}^T$ pairs gives linearly independent columns.  Similarly, the $E_{\psi}$ and $E_{\bar\psi}{}^T$ together can be regarded as comprising a row of $2N$ adjoint Dirac spinors residing in the co-space of the column space mentioned.  So \Eq{double} says that an appropriately defined scalar product between spinor column and adjoint spinor row vanishes identically for any combination of the $\epsilon_\nu$ parameters.   Now suppose that the number of ratios $\epsilon_\nu/\epsilon_1$, e.g. $n^2-2$ for $SU(n)$, is no smaller than $2N$.  Then as the $\epsilon_\nu$s independently sweep over their range of values, the column made up of the $\epsilon_\nu \mathbf{L}_\nu \psi$ and $\epsilon_\nu \mathbf{L}_\nu{}^T  \bar\psi{}^T$ spans the whole $2N$ dimensional space.  How then can the scalar product in \Eq{double} vanish consistently?  Obviously only if the row composed of the adjoint spinors $E_{\psi}$ and $E_{\bar\psi}{}^T$ vanishes identically.   According to \Eq{nonvariation} this means that all field equations for the spinor fields are satisfied.  

Of course the above establishes the field equations only for representations with limited dimensions.  In $SU(n)$, for example, the fundamental representation is of dimension $n$, and $2n$ is less than $n^2-2$ for all $n\geq 3$.  Thus for $SU(3), SU(4), \dots$ the field equations in the fundamental representation of the fields can be obtained by the method just described.  What of higher order representations?  All experience suggests that one obtains the correct field equations for the spinor fields by just replacing the fundamental representation's $\mathbf{L}_\nu$s by the corresponding matrices of the higher order representation, and introducing spinor multiplets of the correct order. What of the lower order unitary groups?  The case $U(1)$ has actually been dealt with already in Sec.~\ref{sec:U(1)}; the case of $SU(2)$ remains outstanding.  It is clear that the method here described should be applicable to many physical systems deriving from the action (\ref{SU(n)}) and subject to many practically interesting symmetry groups.  Again, it is clear that the LAP does not enter at any point. 

\section{Summary}
\label{sec:sum} 
Traditionally for any classical field theory (defined by an action) the least action principle (LAP) is used to derive the field equations.  We have argued here, on the basis of three detailed examples, that it may be possible to get by without the LAP by beginning with the action and then exploiting only the symmetries pertaining to it and energy-momentum conservation.  Conservation laws are here taken to be experimental facts.  Our examples all involve mutually interacting fields; they include a set of most generally interacting real scalar fields in curved space-time, scalar electrodynamics in flat space-time, and a Dirac multiplet in interaction with a nonabelian gauge field in the absence of gravitation.  We have been able to derive the field equations of all field components by using only the action, some of its symmetries and the conservation of energy-momentum.

There are, of course, disadvantages to the latter method.  The strategies for the solution vary from case to case, in contrast to the very standardized procedure for implementing the LAP.  Further, the computations required by the symmetries-based procedure tend to be intricate.  Thus, in practice, the latter method will not replace the LAP approach.   However, provided it is a generally successful approach, the latter method shows that as a matter of principle, the LAP is not an obligatory starting point of a classical field theory.    For this alternative road map to the field equations to be generally applicable an obstacle must be surmounted.  As well illustrated by the scalar fields example, and the spinor-gauge field one, the number of conservation laws may not suffice in every case to determine field equations for all field components.  That is, overly complex physics, e.g. five interacting scalar fields in four dimensions, and on occasion very simple systems, e.g. $SU(2)$ gauge theory with spinors, \textit{may} not be covered by the scheme that sidesteps the use of the LAP.  This issue is still moot, and provides material for further research.

\acknowledgments

We thank Idan Talshir for useful remarks.   This research is supported by the I-CORE Program of the Planning and Budgeting Committee and the Israel Science Foundation (Grant No. 1937/12), as well as by the Israel Science Foundation personal Grant No. 24/12. B.R.M was also supported by  Lady Davis and  Golda Meir postdoctoral fellowships of the Hebrew University.

\end{document}